\documentstyle[prb,aps,twocolumn,epsfig]{revtex}
\newcommand{\cfo}{CuFeO$_2$}
\newcommand{\qtwod}{quasi two-dimensional}
\newcommand{\afm}{antiferromagnet}
\newcommand{\MH}{$M\;vs\;H$}
\newcommand{\MT}{$M\;vs\;T$}

\newcommand\PRB[3]{Phys. Rev. B {\bf {#1}}, {#2} ({#3})}
\newcommand\JPCM[3]{J. Phys.: Condens. Matter {\bf {#1}}, {#2} ({#3})}
\newcommand\PhysB[3]{Physica B {\bf {#1}}, {#2} ({#3})}
\newcommand\JPSJ[3]{J. Phys. Soc. Japan {\bf {#1}}, {#2} ({#3})}
\begin{document}
\draft
\twocolumn[\hsize\textwidth\columnwidth\hsize\csname @twocolumnfalse\endcsname
\title{High Magnetic Field Behaviour of the Triangular Lattice Antiferromagnet, \cfo }
\author{O.~A.~Petrenko$^{1,2}$, G.~Balakrishnan$^2$, M.~R.~Lees$^2$, 
D.~McK~Paul$^2$ and A.~Hoser$^3$}
\address{$^{1}$ ISIS Facility, Rutherford Appleton Laboratory, Chilton, 
Didcot, OX11 0QX, UK \\
$^2$University of Warwick, Department of Physics, Coventry, CV4~7AL, UK \\
$^3$Hahn-Meitner-Institut, Glienicker Stra{\ss}e 100, 14109 Berlin, Germany}
\maketitle
\begin{abstract}
The high magnetic field behaviour of the triangular lattice \afm\ \cfo\ is studied
using single crystal neutron
diffraction measurements in a field of up to 14.5~T and also by magnetisation
measurements in a field of up 
to 12~T. At low temperature, two well-defined first order magnetic phase transitions
are found in this 
range of applied magnetic field ($H\parallel c$): at $H_{c1}=7.6(3)/7.1(3)$~T and
$H_{c2}=13.2(1)/12.7(1)$~T 
when ramping the field up/down. In a field above  $H_{c2}$ the magnetic Bragg peaks
show unusual history dependence. In zero field $T_{N1}=14.2(1)$~K separates a high temperature
paramagnetic and an intermediate incommensurate structure, while $T_{N2}=11.1(3)$~K
divides an incommensurate 
phase from the low-temperature 4-sublattice ground state. The ordering temperature
$T_{N1}$ is found to 
be almost field independent, while $T_{N2}$ decreases noticeably in applied field.
The magnetic phase diagram
is discussed in terms of the interactions between an applied magnetic field and the
highly frustrated magnetic
structure of \cfo . 
\end{abstract}
 ]
\flushbottom
\parskip 0mm
\noindent
\section{Introduction}
     Compounds of the ABO$_2$-type with the delafossite structure provide good
examples of antiferromagnets on a triangular lattice, presenting the opportunity
to study the influence of geometrical
frustration on these magnetic systems~\cite{list}. Unlike the well-studied quasi
one-dimensional ABX$_3$-type
hexagonal antiferromagnets (where A is an alkali metal, B is a bivalent  metal of
the $3d$  group and X is a
halogen)~\cite{Collins_97}, the ABO$_2$ materials are mostly \qtwod\ and are highly
frustrated between
neighbouring triangular layers as well as within a layer. The study of the magnetic
properties of 
ABO$_2$ compounds has been somewhat hampered by the absence of good quality single
crystals.
However, recent progress in crystal growth for one of the members of the
delafossite-family,
\cfo~\cite{Zhao_96}, has revived the interest in this compound and generated a
``second wave" of publications
dealing with its magnetic properties~\cite{Fukuda_98,Galakhov_97,Mitsuda_99,Mitsuda_98}.

     Despite recent efforts, the situation regarding the magnetic properties of \cfo\
is far from clear.
An earlier powder neutron diffraction study~\cite{Mitsuda_91} claimed to observe two
successive magnetic
transitions at $T_{N1}=16$~K and $T_{N2}=10$~K to a collinear spin structure with a
monoclinic unit cell
($\sqrt{7}a \times \sqrt{7}a \times 2c$, five spins in a layer) and an orthorhombic
unit cell ($\sqrt{3}a \times 
2a \times 2c$, four spins in a layer) respectively~\cite{Mekata_92}. All the spins
were found to be parallel to the
$c$-axis and an Ising nature for the exchange interactions has been suggested, although
that would be unusual for
the $^6S$ state of Fe$^{3+}$ ions. Further neutron powder diffraction measurements as
well as the single crystal
susceptibility and M{\" o}ssbauer effect studies~\cite{Mekata_93} 
also showed two successive phase transitions, however it was suggested that the
intermediate temperature (IT) 
phase is a partially disordered phase, where 1/5 of the magnetic moments remain
paramagnetic. The presence of disordered
spins in the IT phase agrees with the heat capacity data~\cite{Takeda_94}, but contradicts
the simple mean-field theory for the Ising antiferromagnet on a triangular lattice which
suggests a partially
disordered phase, where one third of the spins are paramagnetic, rather than one fifth.

     The most interesting effects have been found when \cfo\ is placed in a magnetic field.
As many as five spin-flop like magnetisation anomalies  have been observed for a field applied along the $c$-axis 
(at 8, 13, 22, 26, 42 and 70~T)~\cite{Ajiro_94}, while for $H\perp c$, a phase transition has been found at 24~T. 
Monte Carlo simulation results have suggested that in order for a collinear structure to be stable in zero
field~\cite{Mekata_93} and also undergo several phase transitions in applied field, the second neighbour 
exchange, $J_2$, as well as the third neighbour exchange, $J_3$, have to be unusually strong in comparison 
with the main  nearest-neighbour exchange interaction $J_1$: $J_2/J_1$=0.5 and $J_3/J_1$=0.75.  Although a 
theoretical model with such parameters produces a somewhat similar magnetisation curve, comparable values 
of the exchange interactions $J_1$ and $J_3$ for magnetic moments separated by twice the distance seems to be 
suspicious.  In fact, weak interplane interactions which have been neglected in this study might play
an important role in the formation of the long range magnetic order.

     The exact magnetic structure of the IT phase is still a subject of discussion. Submillimeter wave ESR 
measurements~\cite{Fukuda_98} have discovered an easy-plane antiferromagnetic resonance mode which could 
not be explained in the partially disordered model suggested by previous powder neutron diffraction experiments. 
Recently the first single crystal neutron diffraction study~\cite{Mitsuda_98} has shown that between $T_{N1}$ 
and $T_{N2}$ a quasi-long range amplitude-modulated magnetic structure exists. This structure has an 
incommensurate, temperature dependent, propagation vector $(qq0)$ with $q$ ranging from 0.19 to 0.22. 

\begin{figure}[hp]
\centerline{\psfig{figure=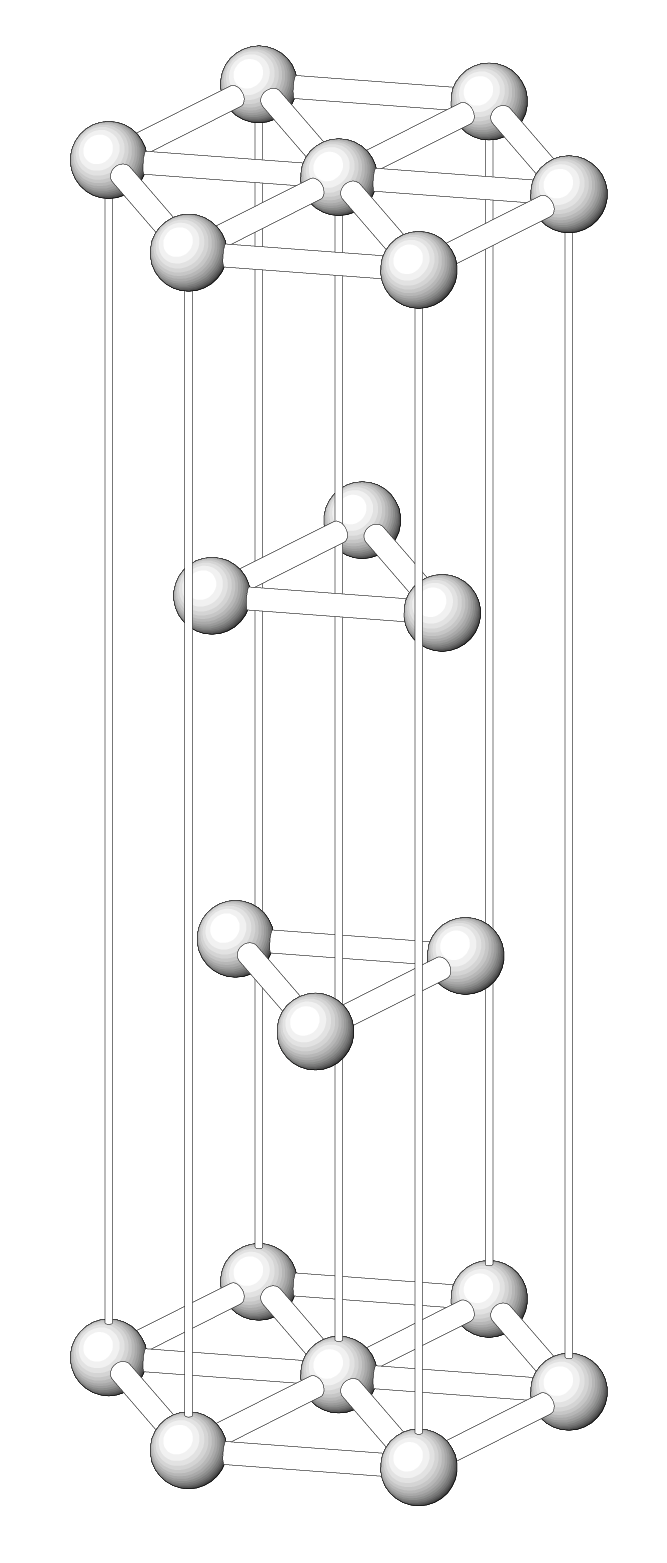,height=9cm,angle=0}}
\caption{Crystal structure of \cfo . Only Fe$^{3+}$ magnetic ions are shown.}
\label{Structure}
\end{figure}

     The crystal structure of \cfo\ belongs to the space group $R\bar{3}m$ with $a$=3.03~\AA\ and $c$=17.09~\AA\
in the hexagonal description. The structure consists of hexagonal  layers of Cu, O and Fe with a stacking
sequence of O$^{2-}$ - Fe$^{3+}$ - O$^{2-}$ - Cu$^{+}$ - O$^{2-}$ - Fe$^{3+}$ - O$^{2-}$  along the
$c$ axis, where the triangular lattice of magnetic Fe$^{3+}$ ions are separated by nonmagnetic 
ionic layers of Cu$^+$ and O$^{2-}$, see Fig.~\ref{Structure}. Therefore, a \qtwod\ character for the magnetic
exchange interactions is the most likely situation, however no actual measurements of the exchange parameters have
been reported so far and the extent of the low-dimensionality of the system remains unknown.

     In this paper we report on single crystal heat capacity, magnetisation and neutron diffraction measurements.
Our main experimental efforts were focused on the properties of the low-temperature (LT) collinear phase in a
magnetic field applied either parallel or perpendicular to the $c$-axis.
We find that contrary to previous suggestions~\cite{Mitsuda_99,Ajiro_94} a
field-induced phase transition from a collinear 4-sublattice state to a
5-sublattice state occurs only when an applied magnetic field exceeds the
value of the second critical field, $H_{c2}\approx 13.2$~T, rather than the
first one, $H_{c1}\approx 7.6$~T. The exact nature of the phase transition
at $H_{c2}$, however, remains unclear.

\section{Experimental Details}
     Polycrystalline \cfo\ was synthesised starting from stoichiometric quantities of the oxides CuO and Fe$_2$O$_3$. 
A mixture of the oxides was first reacted at 800$^\circ$C for 15 hours in air followed by 24 hours in flowing
nitrogen gas with intermediate grindings. Rods of diameter 8-10~mm and length 8-10~cm were then isostatically
pressed from the reacted powder, and sintered at 950$^\circ$C in flowing nitrogen gas. Single crystals of \cfo\
were grown by the floating zone method using a four mirror image furnace (model CSI FZ-T-10000-H-IV-VPS).
The rods were rotated at 30~rpm and the growth was carried out at 1~mm/h, in 1 atmosphere of argon gas
with a flow of 0.3~l/min. The crystals produced had facets and the growth axis was found to be inclined to the
$c$-axis at approximately 10 degrees.

     The specific heat measurements in zero field were made in the temperature range 2 to 20~K using a 
standard heat pulse-relaxation method and from 20 to 305~K using an adiabatic technique. The data were
collected while warming the sample.  

     The magnetisation data, \MH\ and \MT , were collected using an Oxford  Instruments vibrating 
sample magnetometer in an applied field of up to 12~T. For both the specific heat and magnetisation measurements 
we have used small (5 to 40 mg) plate-like samples that were cut from the crystal used for neutron scattering
experiments. The absolute accuracy of the magnetisation measurements was better than 1\%. The crystals were aligned
with an accuracy of 1-2$^{\circ}$.

     The neutron scattering measurements were carried out using the E1 triple-axis spectrometer at the Berlin Neutron 
Scattering Center, HMI, Germany. The spectrometer was used in the double-axis mode, with a pyrolytic graphite
(002) monochromator providing a neutron wavelength of 2.42~\AA . A pyrolytic graphite filter was installed in the
incident beam to remove higher order contamination. 

     The single crystal sample of almost cylindrical shape (13.5 mm in length and 6.5 mm in diameter) was aligned 
in a vertical field cryomagnet VM-1 in two different orientations. In the first case the horizontal scattering
plane contained the $(hk0)$ reflections and in the second case the $(hhl)$ reflections. The horizontal
collimation was 40'--80'--40', typical counting times were 1 to 5 seconds at each point.

\section{Experimental Results} 
\subsection{Low temperature specific heat data} 
\begin{figure}[hp]
\centerline{\psfig{figure=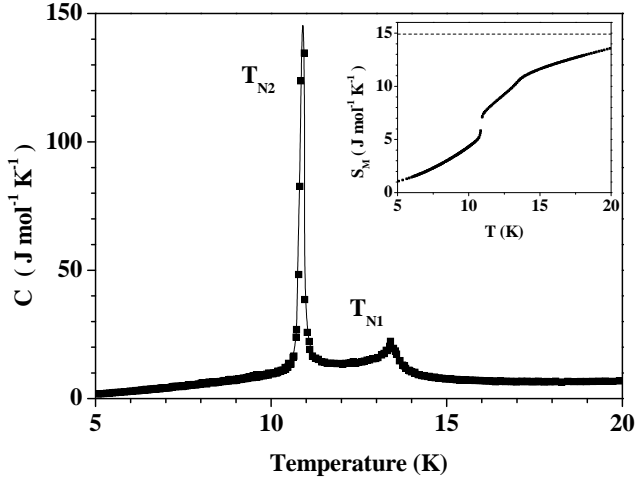,width=\columnwidth,angle=0}}
\caption{Low-temperature heat capacity of \cfo\ single crystal. 
Solid line is a guide for the eye.
The inset shows the temperature dependence of the magnetic entropy, calculated as $S_M(T)=\int\limits_{0}^{T}C/T dT$.
The dashed line indicates the maximum magnetic entropy $R \ln(2{\bf S}+1)=14.90$~J~mol$^{-1}$K$^{-1}$ for ${\bf S}$=5/2.}
\label{C_T}
\end{figure}

      The low-temperature specific heat measurements of the \cfo\ single crystal (see Fig.~\ref{C_T}) show
a remarkable difference between the two magnetic phase transitions. The first transition into the IT phase
is marked by a broad peak in the specific heat at a temperature of around 14~K. The second transition into the
LT phase is accompanied by a much sharper, more intense peak at $T\approx 11$~K. 

Previous thermal analysis measurements~\cite{Takeda_94} claimed that the
magnetic entropy associated with the anomaly in the heat capacity around
$T_{N2}$ is attributable to the disordering of one-fifth of the Fe$^{3+}$ spins.
However, this conclusion seems to be based on an unreliable estimate of the
heat capacity baseline (see Fig.~3 in Ref.~\cite{Takeda_94}). The data
presented here suggests that nearly one third of the Fe$^{3+}$ spins are
disordered by the time the temperature is increased to $T_{N2}$. This
observation is in good agreement with the value expected from the simple
mean-field theory for an Ising antiferromagnet on a triangular lattice.

      Our high-temperature specific heat data indicate that \cfo\ has a relatively high Debye temperature,
$\theta_D\approx600$~K. Therefore, the phonon contribution to the specific heat at low temperatures
is negligibly small
compared to the magnetic contribution. The presence of short range magnetic correlations dominates the
heat capacity at all temperatures up to 40~K, well above the magnetic ordering temperature. 
In fact only about three quarters of the magnetic entropy 
is recovered at $T<T_{N1}$, as shown on the inset to Fig.~\ref{C_T}, while the remaining quarter of the
entropy is spread over  $T>T_{N1}$. The lattice contribution to the entropy, which does not exceed 2\% of the
total entropy at all temperatures shown on Fig.~\ref{C_T}, has been neglected in these calculations.

\subsection{Magnetisation data} 
\begin{figure}[hp]
\centerline{\psfig{figure=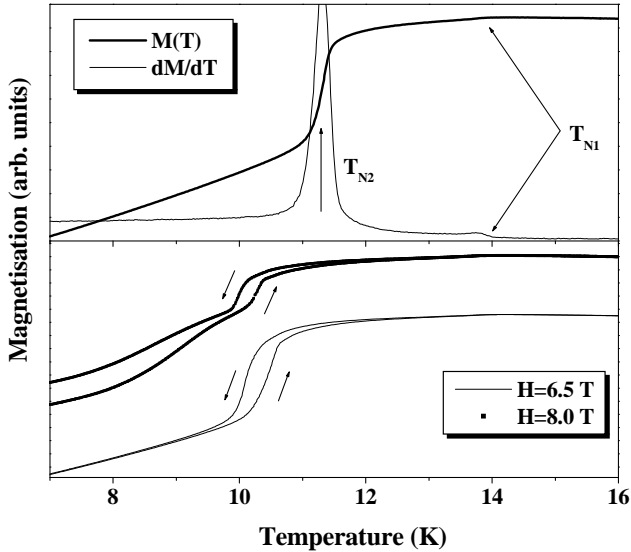,width=\columnwidth,angle=0}}
\caption{Temperature dependence of the magnetisation $M_{H\parallel c}$ (in arbitrary units) measured for a
\cfo\ single crystal at: $H=0.05$~T while warming the sample (upper panel); $H=6.5$~T and $H=8.0$~T while
warming and cooling the sample (lower panel).}
\label{M_T}
\end{figure}

     The first order nature of the lower temperature phase transition is clear from the temperature
dependence of the magnetisation data. $M_{H\parallel c}$ exhibits a dramatic jump at $T_{N2}$, while a small 
kink at $T_{N1}$ became obvious only after differentiation of the data~(see Fig.~\ref{M_T}). 

     An interesting feature was noticed when the sample was warmed and then cooled in an applied 
magnetic field. If the field is below a spin-flop phase transition at $H_{c1}$, then the hysteresis 
in the magnetisation is evident only around $T_{N2}$, while in a field above $H_{c1}$ the magnetisation curves
$M(T)$ for increasing and decreasing temperature are significantly different for all temperatures below $T_{N2}$. 
These results are consistent with the neutron diffraction data (see below). 

\begin{figure}[hp]
\centerline{\psfig{figure=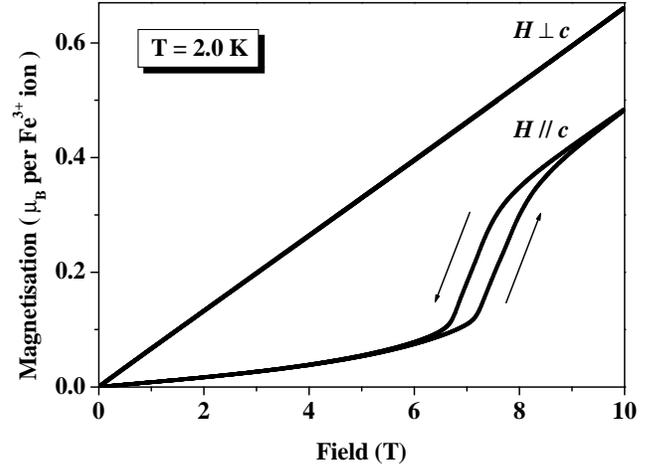,width=\columnwidth,angle=0}}
\caption{Field dependence of the magnetisation at $T=2.0$~K for a \cfo\ single crystal.}
\label{M_H}
\end{figure}

     In a low applied field \cfo\ is magnetically highly anisotropic. This can be seen from the \MH\ measurements for
$H\perp c$ and $H\parallel c$ (Fig.~\ref{M_H}). 
Note that the experimentally observed ratio of the susceptibilities
parallel and  perpendicular to c, $\chi_\parallel/\chi_\perp\approx$0.12 is
inconsistent with the 120$^\circ$ triangular magnetic structure, where the
value of 0.5 is expected at zero temperature.
The magnetic anisotropy
decreases abruptly in a field above the spin-flop phase transition at $H_{c1}=7.6(3)/7.1(3)$~T, however it does
not disappear completely as expected for a simple three-dimensional collinear \afm. Even at $H=10$~T $M_{H\parallel c}$ is still
only 73\% of $M_{H\perp c}$. It would be very interesting to see whether $M_{H\perp c}$
and $M_{H\parallel c}$ do become equal in some higher field (above the second and possibly third transition fields, 
$H_{c2}$ and  $H_{c3}$) or if they remain different, as has been found in the easy-plane ABX$_3$-type 
\afm s~\cite{CsMnBr3_RbMnBr3}.

\subsection{Neutron diffraction data} 
\subsubsection{$(hhl)$ scattering plane, $H\perp c$.} 
     In zero magnetic field the LT magnetic structure is characterised by the appearance
of magnetic Bragg reflections at the positions $(\frac{2n+1}{4}\frac{2n+1}{4}\frac{6m+3}{2})$,
where $n$ and $m$ are integers, in accordance with observations by Mitsuda {\it et al.}~\cite{Mitsuda_98}.
The application of a magnetic field up to 14.5~T along a direction perpendicular to the scattering plane
does not change visibly the position, intensity or width of these peaks -- the 4-sublattice magnetic
structure below $T_{N2}$ remains unchanged by a field applied perpendicular to the $c$-axis.

     At an intermediate temperature, $T_{N2}<T<T_{N1}$, the 4-sublattice magnetic structure
is replaced by an incommensurate phase, in which the magnetic Bragg reflections occur
at positions $(q\;q\;\frac{6m+3}{2})$, where $m$ is an integer and the value of $q$ is temperature dependent
and ranges from 0.19 to 0.225. In the vicinity of $T_{N2}$ and $T_{N1}$ the width of the magnetic
peaks is not resolution limited,
which suggests the absence of true long-range magnetic ordering. However at all temperatures away from the
phase boundaries, that is at 11~K$<T<$~13~K, the width of the magnetic peaks is resolution
limited. It is worth noting, that at $T=10.5$~K, both commensurate and incommensurate peaks are present in the
diffraction pattern. These results are in general agreement with the previous findings of Mitsuda
{\it et al.}~\cite{Mitsuda_98}.

     As with the LT phase, the application of a magnetic field $H\perp c$ does not affect
the characteristics of the IT phase. It only marginally decreases the temperature
of the incommensurate to commensurate phase transition, $T_{N2}$, lowering it to 10~K in 14.5~T.

\subsubsection{$(hk0)$ scattering plane, $H\parallel c$.} 
     In zero magnetic field no purely magnetic reflections have been found in the $(hk0)$ scattering plane.
On the application of a magnetic field above 13~T along the $c$-axis at low temperature, scans in the $(hk0)$
plane of the reciprocal lattice revealed new magnetic Bragg peaks at position $(\frac{n}{5}\frac{m}{5}0)$,
where $n$ and $m$ are nonzero integers and $n+m$ is even. This observation suggests that the magnetic field
induces a transition from a purely antiferromagnetic ordering in zero field to an ordering with a ferromagnetic
component along the $c$-axis.

\begin{figure}[hp]
\centerline{\psfig{figure=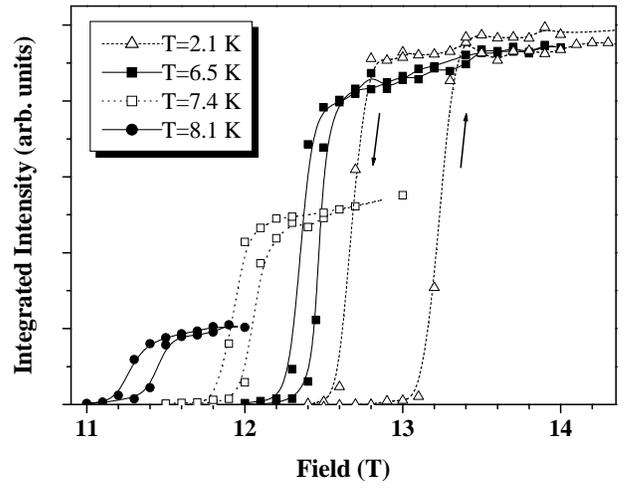,width=\columnwidth,angle=0}}
\caption{Integrated intensity of the $(\frac{1}{5}\frac{1}{5}0)$ peak of \cfo\ in applied magnetic field, 
$H\parallel c$, at different temperatures.}
\label{1/51/50}
\end{figure}

The transition to the high-field phase is accompanied by a significant hysteresis, of about 0.5~T.
The transition field and also the hysteresis width gradually decrease as the temperature increases, though even at
$T=8.1$~K, a temperature relatively close to the phase transition point
$T_{N2}$, the hysteresis amounts to only 0.2~T. We have found no
indication of any short range order within the high field induced magnetic
phase -- the widths of the new magnetic peaks are resolution limited. The
field dependence of the integrated intensity of one of the peaks,
$(\frac{1}{5}\frac{1}{5}0)$, is shown on Fig.~\ref{1/51/50}. The rest of
the $(\frac{n}{5}\frac{n}{5}0)$-type peaks behave similarly.

\begin{figure}[hp]
\centerline{\psfig{figure=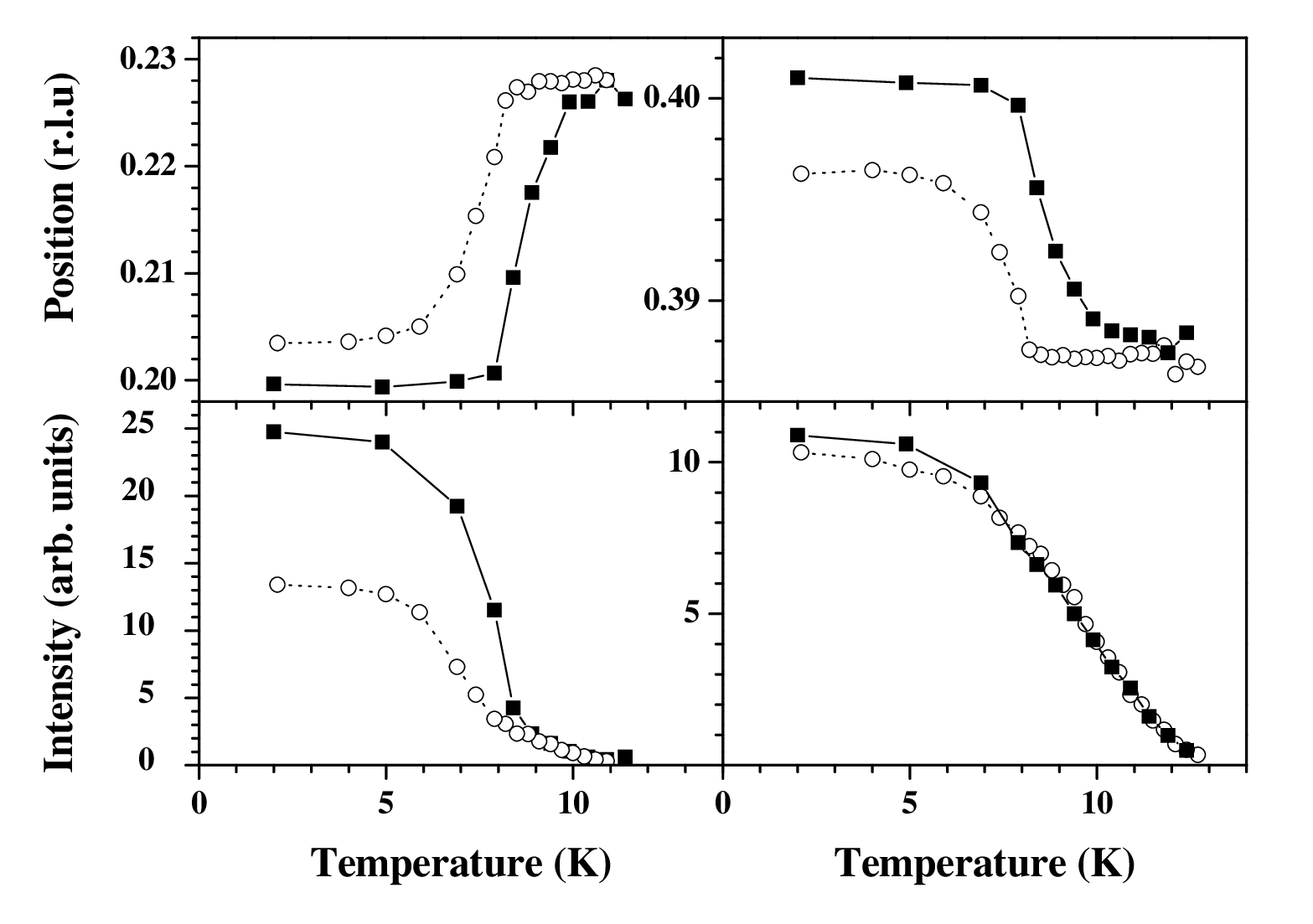,width=\columnwidth,angle=0}}
\caption{Temperature dependence of the position, $n$, (upper panel) and the integrated intensity (lower panel) of two
magnetic peaks of type $(\frac{n}{5}\frac{n}{5}0)$ of \cfo\ in an applied magnetic field of 14.5~T,  $H\parallel c$. Solid and
open symbols correspond to warming and cooling of the sample respectively.}
\label{0204}
\end{figure}

The temperature dependence of the intensity, width and positions of the $(\frac{n}{5}\frac{n}{5}0)$
peaks show remarkable variations in behaviour for different values of $n$. Fig.~\ref{0204} shows the temperature
dependence of the integrated intensity and also of the position in the reciprocal lattice, $n$,
for the two such peaks, with $n\approx 1$ and $n\approx 2$. Initially the sample was cooled
down to 2 K in a zero field, and then the magnetic field was increased to
14.5 T. Warming the sample above $T_{N2}$ in the magnetic field results in a
shift of the peak positions from $(\frac{1}{5}\frac{1}{5}0)$ to $(0.225, 0.225, 0)$ and  from
$(\frac{2}{5}\frac{2}{5}0)$ to $(0.385, 0.385, 0)$.
The higher temperature peaks are still located along the $(hh0)$ direction. When the sample was cooled back
down to 2~K in an applied magnetic field, the first peak with $n\approx 1$ and originally in the
$(\frac{1}{5}\frac{1}{5}0)$ position has only regained approximately half its full integrated intensity
and has remained in the incommensurate position (0.203(1),0.203(1),0). For the second reflection with
$n\approx 2$ the loss of integrated intensity was significantly smaller, while the change of its position from
commensurate $(\frac{2}{5}\frac{2}{5}0)$ to incommensurate (0.396(1),0.396(1),0) was still evident
(see Fig.~\ref{0204}). Other peaks of $(\frac{n}{5}\frac{m}{5}0)$-type demonstrate similar history dependence,
e.g. as a result of thermal cycling in a field, the $(\frac{3}{5}\frac{3}{5}0)$ peak shifts its position to
$(0.604, 0.604, 0)$ and the $(\frac{4}{5}\frac{4}{5}0)$ peak to $(0.795, 0.795, 0)$.

\begin{figure}[hp]
\centerline{\psfig{figure=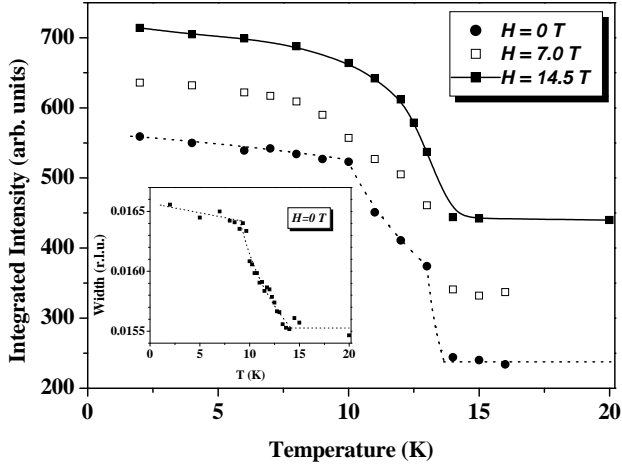,width=\columnwidth,angle=0}}
\caption{Temperature dependence of the intensity of $(110)$ peak in \cfo\ in applied magnetic field, 
$H\parallel c$. For clarity, the curves for $H=7$~T and $H=14.5$~T have been offset by 100 and 200 respectively.
The inset shows the temperature dependence of the linewidth of this peak in zero field.}
\label{110}
\end{figure}

     At temperatures below the magnetic ordering an unusual behaviour of the $(110)$ peak has been observed.
In zero magnetic field the intensity of this peak increases in the IT phase and then remains almost
constant in the LT phase. When the crystal was aligned with the $c$-axis in the horizontal plane the overall
gain in intensity amounted to nearly 10\% of the nuclear intensity at $T>T_{N1}$. A much more pronounced
gain in intensity of the same peak has been seen when the crystal was aligned with the $c$-axis vertical
(see Fig.~\ref{110}). In this case the integrated intensity at $T=2$~K is more than doubled compared to the
intensity at $T>T_{N1}$. An abrupt change in slope in the intensity versus temperature curve is clearly visible
at $T=T_{N2}$. In an applied magnetic field this kink gradually disappears, so that at $H=14.5$~T no anomaly
associated with $T_{N2}$ is visible.

     Mitsuda {\it et al.} have also detected a significant increase of the $(110)$ peak intensity and have
attributed this to a dramatic variation of the crystal mosaic, based on the fact that $(110)$ is not present in
the neutron powder data~\cite{Mitsuda_98}. We note that the observed change in width of the $(110)$ peak
is only 7\% and that the other possible explanations of this effect could be either multiple scattering involving
magnetic peaks or a truly magnetic nature of the $(100)$ peak. The former suggestion is complicated
by the fact that the strongest observed magnetic peak is only half the magnitude of the gain in intensity of
the $(110)$ peak. The latter suggestion has to be dismissed on the basis of a polarised neutron
scattering experiment. The results of this experiment show that for the $(110)$ reflection the ratio of the
neutron scattering intensities with the spin flipper on and off remains practically unchanged when passing
through the phase transition temperatures, $T_{N1}$ and $T_{N2}$. Therefore further investigations of the nature of the low-temperature behaviour of the $(110)$ peak are required.

\section{Discussion and Conclusions}
     We have noted in the Introduction that it is fairly unusual for the $^6S$ state of Fe$^{3+}$ ions to develop
exchange interactions of the Ising-type, suggested by the previous investigations~\cite{Mekata_92}. On the
other hand, it is also hard to imagine how a magnetic system with more degrees of freedom (Heisenberg or XY-type)
could demonstrate as many as six first order-type phase transition in an applied magnetic field. The most likely answer
to this puzzle is the presence in \cfo\ of sufficiently strong competition between the easy-axis anisotropy and the
exchange interactions. Therefore it is essential to describe the magnetic phase diagram of \cfo\ as accurately as possible
and to compare it to similar systems in order to place \cfo\ into the general picture of highly frustrated magnets.

\begin{figure}[hp]
\centerline{\psfig{figure=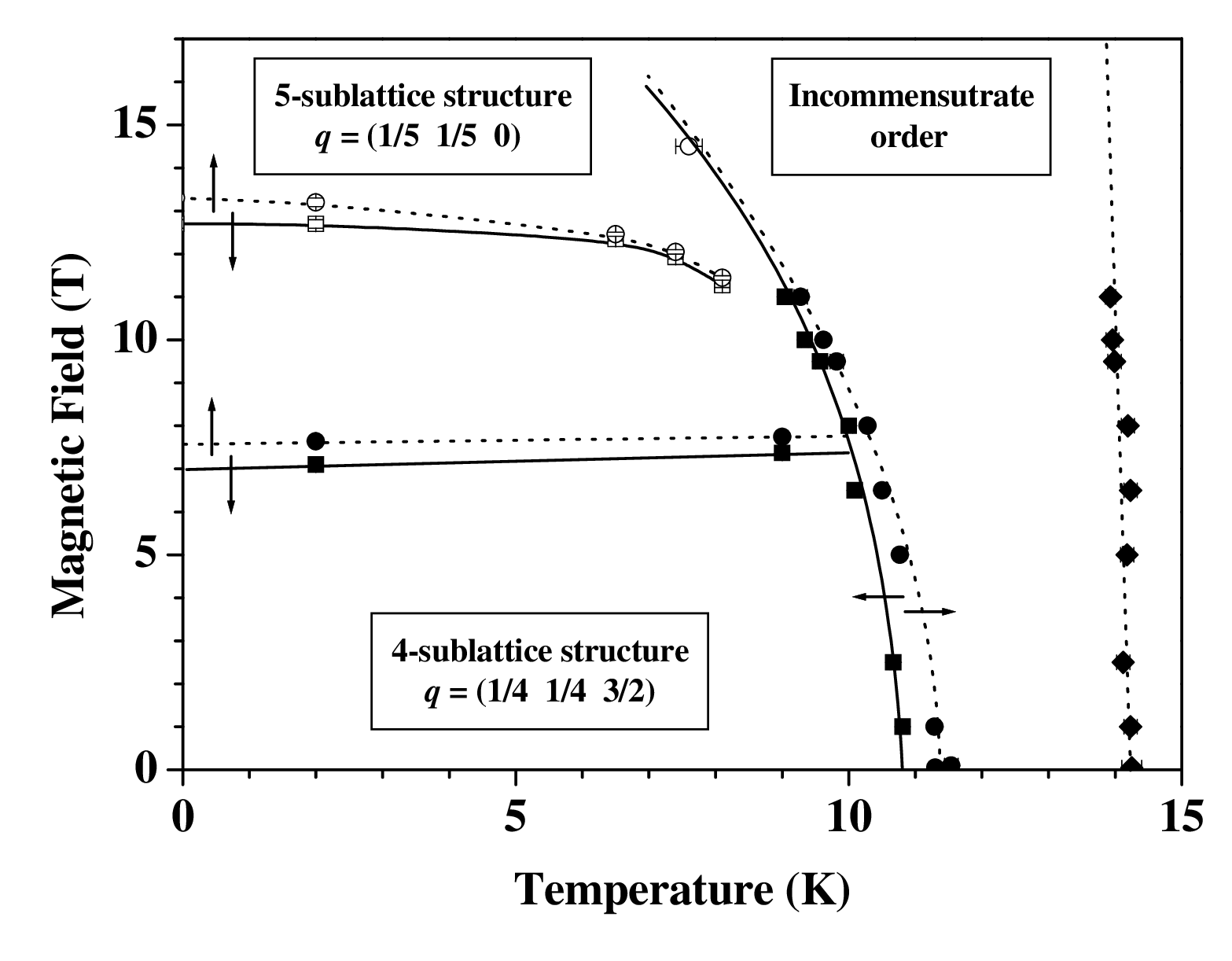,width=\columnwidth,angle=0}}
\caption{Magnetic phase diagram of \cfo , $H\parallel c$, as obtained from magnetisation (solid symbols) and
neutron diffraction (open symbols) data. Solid and dashed lines  correspond to decreasing and increasing
temperature or magnetic field respectively. Only the transition from paramagnetic to an incommensurate phase
is of second order. The rest of the transitions are of first order with a well pronounced hysteresis.}
\label{PhD}
\end{figure}

An overall phase diagram of \cfo\ for a magnetic field applied along the
hexagonal axis is shown in Fig.~\ref{PhD}. It contains a high temperature
paramagnetic phase, an intermediate temperature incommensurate magnetic
phase and also three different commensurate magnetic phases at low
temperature. Contrary to previous results \cite{Mitsuda_98}, we find no
evidence of shorter-range magnetic order in the IT phase at all
temperatures except very close to the phase boundaries. At low temperature
a magnetic field above $H_{c2}\approx 13.2$~T induces a magnetic state
characterized by an ordering wavevector ${\bf
q}=(\frac{1}{5}\frac{1}{5}0)$.

The nature of the lower-field phase transition at $H_{c2}\approx 7.6$~T is
less clear. In order to judge whether it involves Heisenberg degrees of
freedom or corresponds to an Ising type rearrangement of the ${\bf
q}=(\frac{1}{4}\frac{1}{4}\frac{3}{2})$-structure, we would need to make a
survey of all the magnetic peak intensities. To do this it would be
necessary to measure scattering in the $(hhl)$ plane while the magnetic
field is applied along the $c$-axis, which means that the field must be
applied in the horizontal plane. No such facilities are currently
available.

Surprisingly enough the magnetic phase diagram of \cfo\ more closely
resembles the $H-T$ phase diagrams of the rare-earth intermetallic
compounds~\cite{Gignoux_93}, rather than those of the ABX$_3$-type stacked triangular Ising
antiferromagnets~\cite{Gaulin_94}, such as CsCoBr$_3$ and CsCoCl$_3$. In particular, an incommensurate magnetic
ordering occurring in rare-earth intermetallic compounds at $T_{N1}$ exhibits at some lower temperature a transition
to a simpler structure with a shorter period. This happens because in the case of axially anisotropic systems, an
incommensurate or long-period commensurate modulated structure is unstable at 0~K~\cite{Gignoux_93}. At lower
temperature an applied magnetic field lowers the free energy of the incommensurate phases more rapidly than that of
the simple commensurate phases, therefore a reappearance of the incommensurability is often observed in these
compounds in a magnetic field.

     Splitting of $T_N$ is also a common feature for ABX$_3$-type stacked triangular Heisenberg antiferromagnets with
an easy-axis type of anisotropy~\cite{Collins_97}, however their behaviour in an applied magnetic field are profoundly
different to that of \cfo. Their magnetisation curves demonstrate only one spin-flop transition at a critical field $H_c$ for
a field applied along the hexagonal axis and the absence of any transitions for the field applied perpendicular to this axis.
In addition, in ABX$_3$-type systems the application of a magnetic field results in a decrease of the temperature difference
between two phase transition, $T_{N1}-T_{N2}$, while for \cfo\ the effect is the opposite -- an applied magnetic field stabilises
an intermediate incommensurate structure.

In high magnetic field both magnetisation and neutron diffraction measurements show an unusual history dependence.
Cooling the sample in a high magnetic field results in a higher net magnetisation and also in the locking of the magnetic
peaks in the incommensurate positions.

We should also mention the possibility of a high sensitivity of the magnetic properties of \cfo\ to the presence of chemical
impurities~\cite{Ajiro_95} and to the sample preparation technique. The results of Mitsuda {\it et al.}~\cite{Mitsuda_98}
appear to be somewhat sample dependent. Our own magnetisation measurements have revealed a significant
ferromagnetic component in the net magnetisation of those single crystals that have been obtained
from our early attempts of crystal growth, before the growth conditions were optimised. The results of measurements of
magnetisation, heat capacity and neutron diffraction on optimally prepared crystals, however, are self-consistent and also 
sample independent. 

In conclusion, we have presented the results of neutron elastic scattering measurements, heat capacity and
magnetisation studies on a single crystal of \cfo , a \qtwod\ \afm\ on a triangular lattice. These measurements
have allowed a detailed study of the unusual low temperature magnetic properties of \cfo\ in an applied magnetic field
of up to 14.5~T. 

\section{Acknowledgements}
This work was supported  by the  Engineering  and  Physical  Sciences Research Council.
O.~P. acknowledges financial support by the Hahn-Meitner-Institut, Berlin.

\vspace*{15mm}


\begin{thebibliography}{99}
\bibitem{list} The list of ABO$_2$-type compounds for which the magnetic
properties have been studied to different degrees, include ACrO$_2$
(A=Li, Na, K, Cu, Ag and Pd), NaTiO$_2$, LiNiO$_2$, PdCoO$_2$, \cfo,
AgNiO$_2$, CuNdO$_2$ and CuMnO$_2$.
\bibitem{Collins_97}    M.~F.~Collins and O.~A.~Petrenko, Can.~J.~Phys.  {\bf 75}, 605 (1997).
\bibitem{Zhao_96}       T.-R.~Zhao, M.~Hasegawa and H.~Takei, J. Cryst. Growth {\bf 166}, 408 (1996); 
                                  {\bf 154}, 322 (1995).
\bibitem{Fukuda_98}   T.~Fukuda, H.~Nojiri, M.~Motokawa, T.~Asano, M.~Mekata and Y.~Ajiro,
                                 \PhysB{246-247}{569}{1998}.
\bibitem{Galakhov_97} V.~R.~Galakhov, A.~I.~Poteryaev, E.~Z.~Kurmaev, V.~I.~Anisimov, St.~Bartkowski,
                                  M.~Neumann, Z.~W.~Lu, B.~M.~Klein and T.-R.~Zhao, \PRB{56}{4584}{1997}.
\bibitem{Mitsuda_99}   S.~Mitsuda, T.~Uno, M.~Mase, H.~Nojiri, K.~Takashi, M.~Motokawa and M.~Arai,
                                 J. Phys. Chem. Solids {\bf 60}, 1249 (1999).
\bibitem{Mitsuda_98}   S.~Mitsuda, N.~Kasahara, T.~Uno and M.~Mase, \JPSJ{67}{4026}{1998}.
\bibitem{Mitsuda_91}   S.~Mitsuda, H.~Yoshizawa, N.~Yaguchi and M.~Mekata, \JPSJ{60}{1885}{1991}.
\bibitem{Mekata_92}   Although reporting apparently the same data as in Ref.\cite{Mitsuda_91}, the authors
                                M.~Mekata {\it et al.} quote a different value for the lower temperature phase transition, 
                                $T_{N2}$=11~K, in J. Mag. and Mag. Mat. {\bf 104-107}, 823 (1992).
\bibitem{Mekata_93}   M.~Mekata, N.~Yaguchi, T.~Takagi, T.~Sugino, S.~Mitsuda, H.~Yoshizawa,
                                 N.~Hosoito and T.~Shinjo, \JPSJ{62}{4474}{1993}.
\bibitem{Takeda_94}   K.~Takeda, K.~Miyake, M.~Hitaka, T.~Kawae, N.~Yaguchi and  M.~Mekata,
                                 \JPSJ{63}{2017}{1994}.
\bibitem{Ajiro_94}      Y.~Ajiro, T.~Asano, T.~Takagi, M.~Mekata, H.~Aruga-Katori and T.~Goto,
                               \PhysB{201}{71}{1994}.
\bibitem{CsMnBr3_RbMnBr3} In CsMnBr$_3$~\cite{Zaliznyak_92} and RbMnBr$_3$~\cite{Abanov_94} an experimentally 
observed anisotropy between the magnetisation when the field is applied along and perpendicular to the $c$-axis
has been attributed to the influence of quantum~\cite{Abarzhi_92} or thermal~\cite{Santini_96} fluctuations,
enhanced by the low-dimensional nature of the system.
\bibitem{Zaliznyak_92} I.~A.~Zaliznyak, Solid State Comm. {\bf 84}, 573 (1992).
\bibitem{Abanov_94} A.~G.~Abanov and O.~A.~Petrenko, \PRB{50}{6271}{1994}.
\bibitem{Abarzhi_92} S.~I.~Abarzhi, A.~N.~Bazhan, L.~A.~Prozorova and I.~A.~Zaliznyak, \JPCM{4}{3307}{1992}.
\bibitem{Santini_96}  P.~Santini, Z.~Domanski, J.~Dong and P.~Erdos, \PRB{54}{6327}{1996}.
\bibitem{Gignoux_93} D.~Gignoux and D.~Schmitt, \PRB{48}{12682}{1993}.
\bibitem{Gaulin_94} B.~D.~Gaulin, in {\it Magnetic Systems with Competing Interactions}, edited by H.~T.~Diep 
                             (World Scientific, Singapore), 286 (1994).
\bibitem{Ajiro_95}  Y.~Ajiro, K.~Hanasaki, T.~Asano, T.~Takagi, M.~Mekata, H.~Aruga-Katori and T.~Goto,
                               \JPSJ{64}{3643}{1995}.
\end{thebibliography}
\end{document}